\documentclass{PoS}

\title{Dark Matter Related to Axion and Axino}

\ShortTitle{Axion and Axino}

\author{\speaker{Jihn E. Kim}\thanks{This work is supported in part by the KRF Grant No. KRF-2005-084-C00001.}\\
        Department of Physics and Astronomy\\
        Seoul National University\\
        Seoul 151-747, Korea\\
        E-mail: \email{jekim@ctp.snu.ac.kr}}

\abstract{I discuss the essential features of the QCD axion: the strong CP
solution and hence its theoretical necessity. I also review the axion and axino effects on astrophysics and cosmology, in particular with emphasis on their role in the dark matter component in the universe.}

\FullConference{Identification of dark matter 2008\\
		 August 18-22, 2008\\
		 Stockholm, Sweden}

\begin{document}

\newcommand{\Dslash}{{D\hskip -0.23cm\slash}}
\newcommand{\dslash}{{\partial\hskip -0.185cm\slash}}

\section{The Strong CP Problem}

Cosmology with cold dark matter (CDM) was the leading candidate which was started from the late 1970s \cite{LeeWein77} and dominated the field for the next 20 years, which has changed completely just before the beginning the new millennium. The current view of the dominant components of the universe is $\Omega_{\rm CDM}\simeq 0.23$ and $\Omega_\Lambda\simeq 0.73$.
The attractive DM candidates at present is the lightest supersymmetric particle (LSP), axion, axino, and gravitino. Here we will review on axion and the related particle axino. Since axion is the cherished son of the strong CP problem, we begin with the discussion on the strong CP problem and neutron electric dipole moment (NEDM).

All the discussion leading to axion started with the discovery of instanton solutions in nonabelian gauge theories, which has led to an intrinsic additional parameter in QCD, $\theta$. In the $\theta$ vacuum, we must consider the P and T (or CP) violating interaction parametrized by $\bar\theta$,
\begin{equation}
{\cal L}= \bar\theta \{F \tilde F\}
\equiv\frac{\bar\theta}{64\pi^2}\epsilon^{\mu\nu\rho\sigma}
F_{\mu\nu}^aF^a_{\rho\sigma}
\end{equation}
where $\bar\theta=\theta_0+\theta_{\rm weak}$ is the final value including the input value ($\theta_0$) defined above the electroweak scale and the effects of the electroweak CP violation ($\theta_{\rm weak}$). $\bar\theta$ is a physical parameter contributing to the NEDM, $d_n$.

\begin{figure}[!h]
\resizebox{0.95\columnwidth}{!}
{\includegraphics{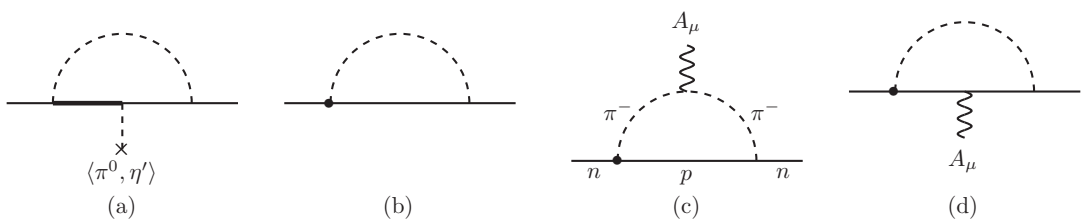}}
\caption{The corrections to the neutron mass and the NMDM. The bullet signals the CP violation.} \label{fig:NEDM}
\end{figure}

In the chiral perturbation theory, the neutron mass and neutron magnetic dipole moment (NMDM) are corrected as shown in Fig. \ref{fig:NEDM}. We start with the vacuum with $\langle \pi^0\rangle= \langle \eta'\rangle=0$ in which CP is conserved. If $\pi^0$ and/or $\eta'$ develops a vacuum expectation value (VEV), then CP is violated.
The loop contribution to the neutron mass term is shown in (a) from which the thick neutron intermediate state is contracted to give (b) where the bullet signals the CP violation via the VEV of $\pi^0$. For the NMDM, we can consider the diagrams shown in (c) and (d). When we redefine the external neutron line to absorb the phase appearing in (b), the phase corresponding to (d) is simultaneously removed. Thus, the remaining unremovable NMDM phase corresponds to (c), which is the NEDM.
From the upper bound of $|d_n|<3\times 10^{-26}e$cm \cite{NEDM}, we obtain a bound on $\bar\theta$, $|\bar\theta|<10^{-11}$ \cite{KimCarosi}.

This small parameter problem is a naturalness problem called the strong CP problem, $\lq\lq$Why is this $\bar\theta$ so small?" Let us classify the known strong CP solutions in three categories:
$$
\rm 1.~ Calculable~ \bar\theta, \quad 2.~ Massless~ up~ quark,
\quad 3.~Axion.
$$
In the calculable $\bar\theta$ models, the Nelson-Barr type CP violation \cite{NBarr} is mostly discussed since it is designed to allow the Kobayashi-Maskawa type weak CP violation at the electroweak scale. It introduces extra heavy quark fields and the interactions beyond the standard model. The scheme is designed such that at low energy the Yukawa couplings are real, which is needed anyway from the beginning to set $\theta_0=0$. This solution is possible with the specific forms for the couplings and their phases and in addition the assumption on VEVs of Higgs doublets \cite{NBarr}.

The second solution is the massless up quark possibility. Even though we will exclude this possibility in the end, let us show the chiral transformation property in detail because exactly this property was the beginning of the invention of the Peccei-Quinn (PQ) symmetry. Suppose that we chiral-transform a quark as $q\to
e^{i\gamma_5\alpha}q$. Then, the QCD Lagrangian changes as
\begin{equation}
\int d^4x [-m_q \bar qq-\bar\theta  \{F \tilde F\} ]\to
\quad\int d^4x [-m_q \bar qe^{2i\gamma_5\alpha}q
-(\bar\theta-2\alpha) \{F \tilde F\}]\label{chiraltr}
\end{equation}
If $m_q=0$, it is equivalent to changing $\bar\theta\to
\bar\theta-2\alpha$. Thus, there exists a shift symmetry
$\bar\theta\to \bar\theta-2\alpha$. In this case, $\bar\theta$ is not physical, and hence there is no strong CP problem if the lightest quark (i. e. the up quark) is massless. However, the recent compilation by Manohar and Sachrajda on the light quark masses in the Particle Data book \cite{Manohar}, $m_u=3\mp 1$ MeV and $m_d=6\pm 1.5$ MeV, is convincing enough to rule out the massless up quark possibility.

\section{Axions, Stars, and the Universe}

\noindent {\bf Axions:} Peccei and Quinn (PQ) tried to mimic the above symmetry  $\bar\theta\to
\bar\theta-2\alpha$ of the massless quark case, by considering the full electroweak theory \cite{PQ77}. They found such a symmetry with an appropriate Higgs potential if $H_u$ is coupled only to up-type quarks and $H_d$ couples only to down-type quarks. Peccei and Quinn succeeded in introducing the $\bar\theta$ shift symmetry, U(1)$_{\rm PQ}$, as in the massless quark case in the electroweak theory. But unlike the massless quark case, here $\bar\theta$ is physical. Weinberg and Wilczek noted that the global symmetry U(1)$_{\rm PQ}$ is spontaneously broken and the resulting Goldstone boson {\it axion} is almost massless \cite{WeinWil}. The axion potential depends on $-\cos\bar\theta$ where $a=\bar\theta F_a$. Since it is proportional to $-\cos\bar\theta$, the vacuum chooses $\bar\theta=0$ as the minimum of the potential. Thus, the axion solution of the strong CP problem is a kind of the cosmological solution.

Nowadays, cosmologically considered axions are considered to be very light, which arises from the phase of SU(2)$\times$U(1) singlet scalar field $\sigma$. The simplest case is the Kim-Shifman-Vainstein-Zakharov (KSVZ) axion model \cite{KSVZ} which incorporates a heavy quark $Q$ with the following coupling and the resulting chiral symmetry
\begin{eqnarray}
{\cal L}=&&-\bar Q_LQ_R\sigma+({\rm h.c.})-V(|\sigma|^2)-\bar\theta
\{F \tilde F\},\nonumber\\
{\cal L}\to &&-\bar Q_Le^{i\gamma_5\alpha} Q_R e^{i\beta}\sigma
+({\rm h.c.})-V(|\sigma|^2)
-(\bar\theta-2\alpha)  \{F \tilde F\}.
\end{eqnarray}
Here, Higgs doublets are neutral under U(1)$_{\rm PQ}$. By coupling $\sigma$ to $H_u$ and $H_d$, one can introduce a PQ symmetry also, not introducing heavy quarks necessarily, and the resulting axion is called the Dine-Fischler-Srednicki-Zhitnitskii (DFSZ) axion
\cite{DFSZ}. In string models, most probably both heavy quarks and Higgs doublets contribute to the $\sigma$ field couplings. The VEV of $\sigma$ is much above the electroweak scale and the axion is a {\it very light axion}. In axion physics, heavy fermions carrying color charges are special. Here, consider an effective theory at the electroweak scale (above the QCD scale), integrating out heavy fields \cite{KimCarosi},
\begin{eqnarray}
{\cal L}_{\theta}&=&\frac12 f_{S}^2\partial^\mu \theta\partial_\mu\theta-\frac14 G_{\mu\nu}^a G^{a\hskip 0.02cm \mu\nu}+(\bar q_L i\Dslash q_L+\bar q_R i\Dslash q_R)
+ c_1(\partial_\mu \theta)\bar q\gamma^\mu\gamma_5 q-\left(\bar q_L~ m~q_R e^{ic_2\theta}+{\rm h.c.}\right)\nonumber\\
&&+c_{3} \frac{\theta}{32\pi^2}G^{a}_{\mu\nu}\tilde G^{a\hskip 0.02cm\mu\nu}\ ({\rm or}\ {\cal L}_{\rm det} ) +c_{\theta\gamma\gamma}\frac{\theta}{32\pi^2}F
^{i}_{\rm em,\mu\nu}\tilde F_{\rm em}^{i\hskip 0.02cm\mu\nu}+{\cal L}_{\rm leptons,\theta}\label{Axionint}
\end{eqnarray}
where $c_1$ term is the derivative coupling respecting the PQ shift symmetry, the $c_2$ term is the phase in the quark mass matrix, the $c_3$ term is the anomalous coupling or the determinental interaction ${\cal L}_{\rm det}$, and $\theta=a/f_S$ with the axion decay constant $f_S$ up to the domain wall number ($f_S=N_{DW}F_a$). ${\cal L}_{\rm leptons,\theta}$ is the axion interaction with leptons. The determinental interaction \cite{tHooft} can be used instead of the $c_3$ term,
\begin{equation}
{\cal L}_{\rm det}=-2^{-1}ic_3\theta(-1)^{N_f}
\frac{e^{-ic_3\theta}}{K^{3N_f-4}} {\rm Det}(q_R\bar q_L)+{\rm h.c.}
\label{detint}
\end{equation}
where we multiplied the overall interaction by $\theta$ in the small $\theta$ region and require the periodicity condition, $c_3\theta=c_3\theta+2\pi$. [The periodicity can be accommodated automatically if we replace $-2^{-1}ic_3\theta$ by 1, but then we must add a constant so that it vanishes at $\theta=0$.] The sign is chosen following Vafa and Witten \cite{VW}. The chiral transformation of quarks show the following reparametrization invariance,
\begin{eqnarray}
\Gamma_{1PI}[a(x), A_\mu^a(x); c_1, c_2, c_3,  m, \Lambda_{\rm QCD}]= \Gamma_{1PI}[a(x), A_\mu^a(x); c_1 -\alpha, c_2-2\alpha, c_3+2\alpha, m, \Lambda_{\rm QCD}].\nonumber
\end{eqnarray}
So, the axion mass depends only on the combination of
$c_2+c_3$. We can convince this also below the ciral symmetry breaking scale. With $u$ and $d$ quarks, we can write an effective Lagrangian below the chiral symmetry breaking scale.
\begin{figure}[!]
\resizebox{1\columnwidth}{!}
{\includegraphics{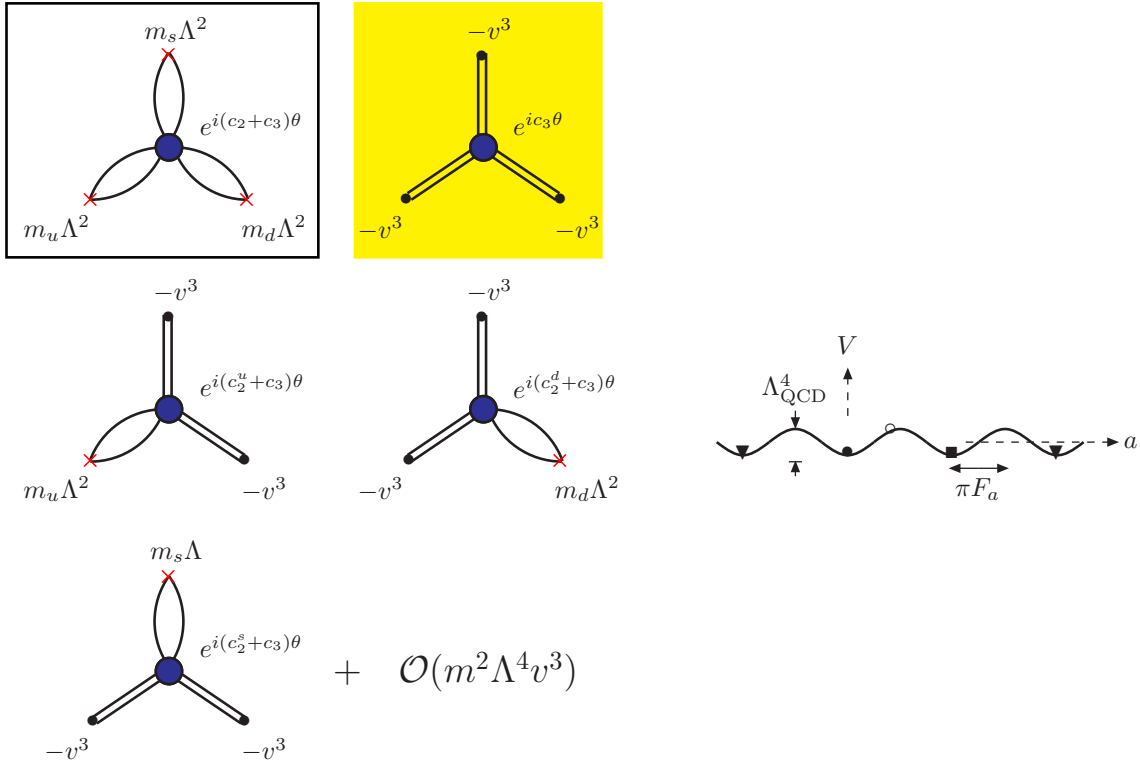}}
\caption{The 't Hooft determinental interaction with possible contractions of light quark lines and the axion potential arising from the second line figures. The yellow shaded one is contributing to the $\eta'$ mass.}\label{fig:DetInt}
\end{figure}
In view of Fig. \ref{fig:DetInt}, we can consider the following $a, \eta', \pi^0$ mass matrix,
\begin{eqnarray}
M^2_{a,\eta',\pi^0}=\left(\begin{array}{ccc}
c^2[\Lambda_{\eta'}^4+2\mu\Lambda_{\rm inst}^3]/F^2 \ \
&-2c[\Lambda_{\eta'}^4+\mu\Lambda_{\rm inst}^3]/f'F& 0\\ &&\\
-2c[\Lambda_{\eta'}^4+\mu\Lambda_{\rm inst}^3]/f'F\ \  & [4\Lambda_{\eta'}^4
 +2\mu\Lambda_{\rm inst}^3+{m_+}v^3  ]/f'^2  & -m_-v^3/ff'\\
&& \\
0 &-m_-v^3/ff' &(m_+v^3+2\mu\Lambda_{\rm inst}^3)/f^2
\end{array}\right)\nonumber
\end{eqnarray}
where $ c=c_2^u+c_2^d+c_3, F=f_S, f=f_\pi,f'=f_{\eta'}$ and $\Lambda_{\eta'}$ and $\Lambda_{\rm inst}$ are QCD parameters, and $m_+=m_u+m_d, m_-=m_d-m_u$, and
$\mu={m_um_d}/{(m_u+m_d)}$. In the limit $f/F, f'/F\ll 1$, we obtain \cite{KimCarosi},
\begin{eqnarray}
m^2_{\pi^0} &\simeq& \frac{m_+v^3+2\mu \Lambda_{\rm inst}^3}{f_{\pi}^2},\
m^2_{\eta'} \simeq \frac{4\Lambda_{\eta'}^4+m_+v^3+2\mu \Lambda_{\rm inst}^3}{f_{\eta'}^2}\\
m^2_a &\simeq&
\frac{c^2}{F^2}\frac{Z}{(1+Z)^2}f_{\pi}^2 m_{\pi^0}^2\left(1+\Delta \right)
{\rm \ \ with \ \ }
\Delta=\frac{m_-^2}{m_+}~\frac{\Lambda_{\rm inst}^3(m_+v^3
+\mu\Lambda_{\rm inst}^3)}{m_{\pi^0}^4 f_\pi^4}.\label{AxmassDelta}
\end{eqnarray}
In this form, the instanton contribution $\Delta$ is included in the axion mass.

Axion is directly related to $\bar\theta$. Its birth was from the PQ symmetry whose spontaneous breaking introduced $a$. Generally, however, we can define $a$ as a pseudoscalar field without potential terms except for the one arising from the gluon anomaly,
$\frac{a}{F_a}\left\{\frac{g^2}{32\pi^2}F^a_{\mu\nu}
\tilde F^{a\mu\nu} \right\}$. Then, we note that this kind of nonrenormalizable term can arise in several ways: string theory and $M$-theory \cite{String}, large extra ($n$) dimensions \cite{extraD}, composite models \cite{composite}, and renormalizable  theories. The axion decay constant is given with the scale defining the model, except in renormalizable models where it is given by the PQ symmetry breaking scale. In any case, the essence of the axion solution (wherever it originates) is that $\langle a\rangle$ seeks $\bar\theta=0$ whatever happened before. The potential arising from the anomaly term after integrating out the gluon field is the axion potential. The height of the potential is $\sim O(\Lambda^4_{\rm QCD})$.
Two important properties of axions are: (i) periodic potential with the period $2\pi F_a$, and (ii) the minima are at $a=0, 2\pi F_a, 4\pi F_a, \cdots$. The cosine form of the potential is usually used with the mass given in (\ref{AxmassDelta}). The axion mass is
$m_a\simeq 0.6 \left(\frac{10^7\rm~ GeV}{F_a}\right){\ \rm eV}.$

Above the electroweak scale, we integrate out heavy
fields. If colored quarks are integrated out, its effect is appearing as the coefficient of the gluon anomaly. If only bosons are integrated out, there is no anomaly term. Thus, we have $c_1=0, c_2=0,$ and $c_3=$ nonzero for the KSVZ axion and $c_1=0, c_2=$ nonzero and $c_3=0$ for the
DFSZ and the PQWW axions. There can be the axion-photon-photon anomalous coupling of the form $a{\bf E} \cdot{\bf B}$. These couplings can be checked in laboratory, astrophysical and cosmological tests. The old laboratory bound of $F_a>10^4$ GeV has been obtained from meson decays ($J/\Psi\to a\gamma, \Upsilon\to a\gamma, K^+\to a\pi^+$), beam dump experiments ($p(e)N\to aX\to \gamma\gamma X, e^+e^-X$), and nuclear de-excitation ($N^*\to Na\to N\gamma\gamma, Ne^+e^-$) \cite{Kim87}.
\\
\vskip -0.3cm
\noindent {\bf Axions from stars:}
We use the axion couplings to $e, p, n,$ and photon to study the core evolution of a star.  The important process is the Primakoff process for which the coupling $c_{a\gamma\gamma}$ is defined as
${\cal L}=-c_{a\gamma\gamma}\frac{a}{F_a}\{F_{\rm em}\tilde F_{\rm em}\}$ with $c_{a\gamma\gamma} =\bar c_{a\gamma\gamma}-1.95$ where --1.95 arises going through the QCD chiral phase transition. The high energy value
$\bar c_{a\gamma\gamma}={\rm Tr} Q^2_{\rm em}|_{E\gg M_Z}$ is obtained from the PQ charges of the colored fermions. In the hot plasma in stars, axions once produced most probably escape the core of the star and take out energy. This contributes to the energy loss mechanism of star and should not dominate the luminocity. Thus, axions from the Sun have been searched by axion helioscopes of Tokyo \cite{Tokyo} and CAST experiments \cite{CAST}. However, the most stringent bound comes from the study of SN1987A \cite{RafTur}. In this case,
the information on the axion--hadron coupling is crucial. So far, the axion--hadron couplings were given for the KSVZ axion \cite{ChangChoi}, but now they are given for the DFSZ axion also \cite{KimCarosi}. The SN1987A gave a strong bound $F_a>0.6\times 10^9$ GeV \cite{RafTur}.

\begin{figure}[!h]
\resizebox{0.8\columnwidth}{!}
{\hskip 3cm \includegraphics{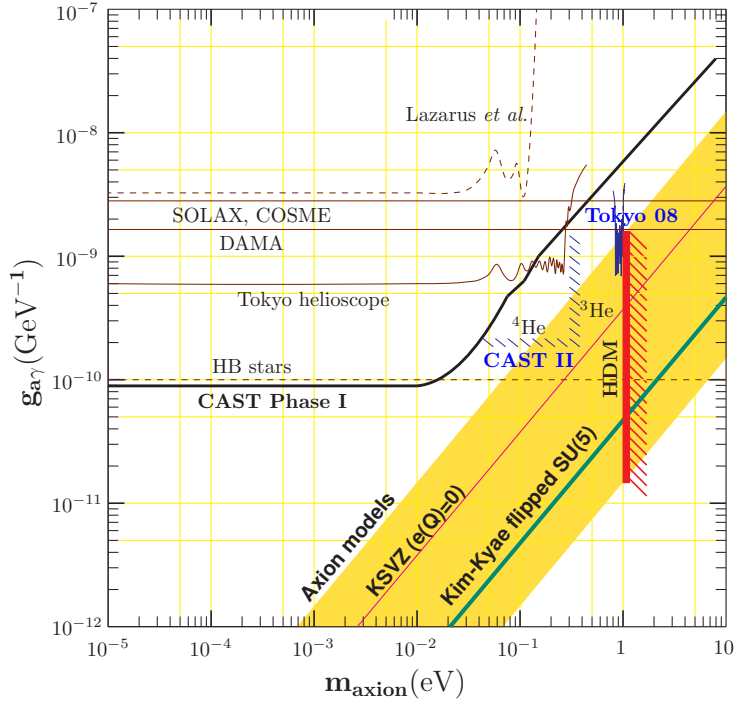} 
}
\caption{The CAST and Tokyo experimental bounds with some theoretical values for $g_{a\gamma}=\alpha_{\rm em}c_{a\gamma\gamma}/2\pi F_a$. For string model, there exists one calculation shown as a green line.\cite{stringcagg}. }\label{fig:CASTexp}
\end{figure}
Laboratory experiments can perform more than just the energy loss mechanism in the core of a star. The early Tokyo experiment could not give a more stringent bound than the supernova limit, but the CAST could compete with the supernova bound. The Tokyo and CAST results are shown in Fig. \ref{fig:CASTexp} together with other theoretical predictions.
\\
\vskip -0.3cm
\noindent{\bf Axions in the universe:} The potential of the very light axion is of almost
flat. Therefore, a chosen vacuum point stays there for a long time, and starts to oscillate when the Hubble time $H^{-1}$ is comparable to the oscillation period (the inverse axion mass), $H<m_a$.  This occurs when the temperature of the universe is about 1 GeV \cite{PWW}. Since the reheating temperature is $T_{\rm RH}<10^9$ GeV or $10^7$ GeV in some models \cite{EKN}, here we will not worry about the domain wall(DW) problem any more.

\begin{figure}[!h]
\resizebox{1\columnwidth}{!}
{\includegraphics{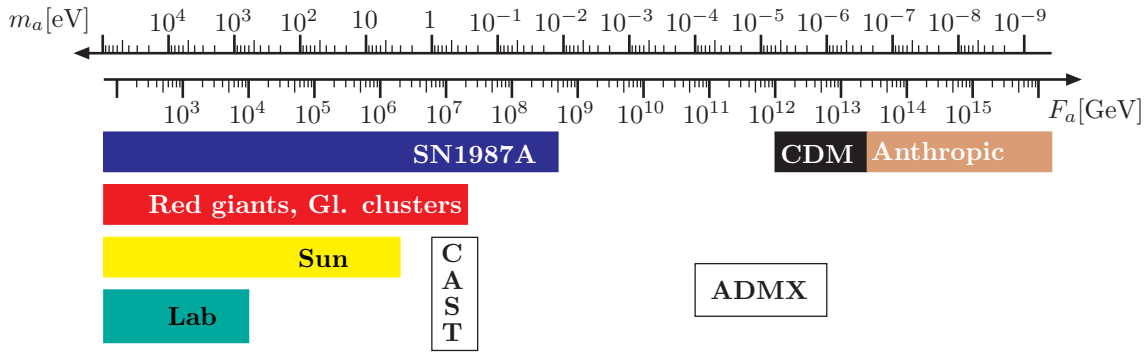}}
\caption{The axion window is not completely closed.  The anthropic region is always allowed. }\label{fig:AxionWindow}
\end{figure}

Since the first cosmological study \cite{PWW}, there appeared a few changes: the values of the light quark masses, the axion CDM energy fraction in the universe, and the QCD phase transition \cite{QCDphase}. In Ref. \cite{Bae08}, these are included and a new overshoot factor is also taken into account. The axion is created at $T=F_a$, but the universe $\langle a\rangle$ does not begin to roll until $H=m_a$, i.e. at $T=0.92$ GeV. From then, the classical field  $\langle a\rangle$  starts to oscillate.
\begin{figure}[!]
\resizebox{1\columnwidth}{!}
{\includegraphics{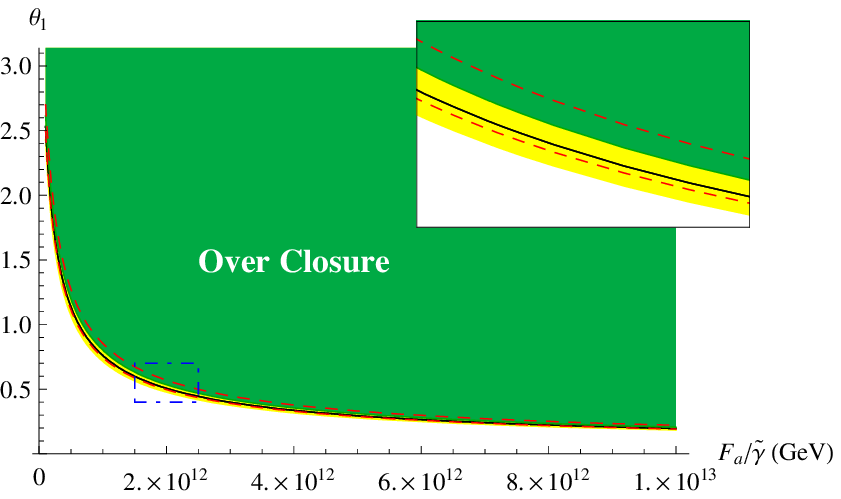}\hskip 1cm \includegraphics{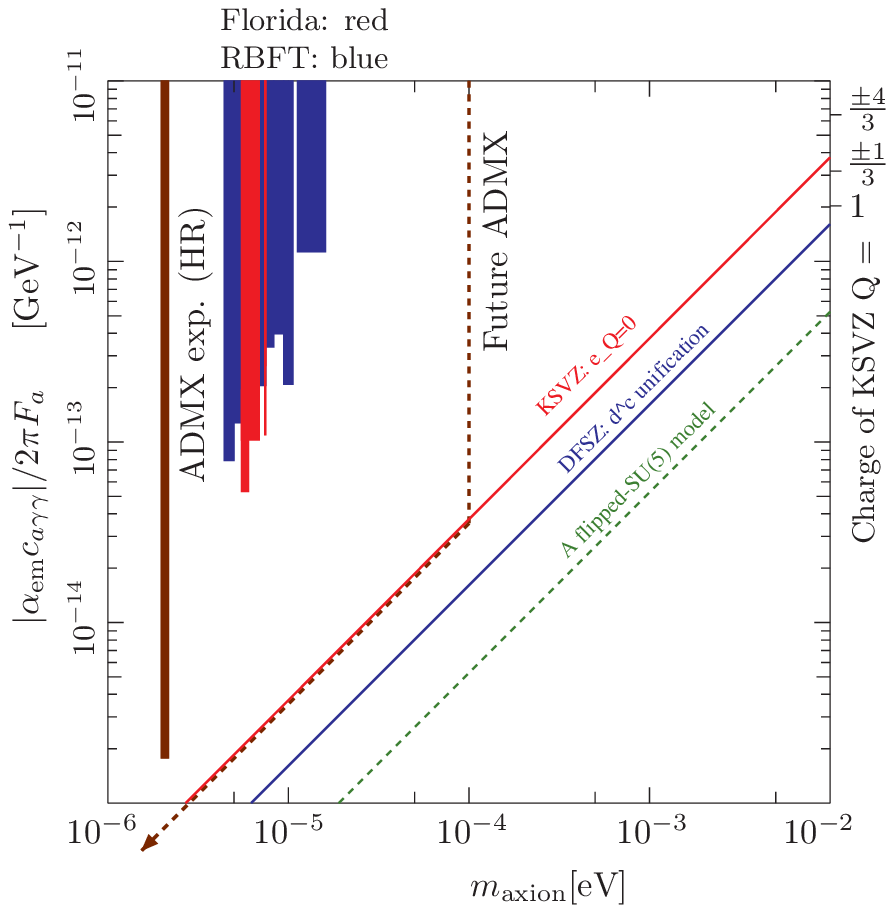}}
\caption{The allowed regions of $F_a$ and the initial misalignment angle $\theta_1$, and the current $F_a$ bound from the cavity experiments. }\label{fig:CosmicAxion}
\end{figure}
This is shown in the left figure of Fig. \ref{fig:CosmicAxion}, where we note that $F_a$ must be less than $10^{12-13}$ GeV depending on the initial misalignment angle. On the right hand side, we compare this prediction of the cosmic axion energy density with the cavity axion search experiments. Summarizing the astro and cosmological constraints, we customarily take the axion $F_a$ window as
$
10^{9}\ {\rm GeV} \le F_a\le 10^{12}\ \rm GeV.
$
 In Fig. \ref{fig:AxionWindow}, we show the open axion window together with the ongoing axion search experiments which is reviewed by van Bibber's talk \cite{vanBibber}.

But there is an anthropic argument even beyond
$F_a> 10^{12}$ GeV, which became popular in recent years. The early anthropic argument on axion was given by Pi and Linde \cite{Pi} and the recent more refined version is given by Tegmark {\it et al.} \cite{Tegmark06}. The homogeneous axion field value (with $a\to -a$ symmetry) right after inflation can take any value between 0 and $\pi F_a$ or $\bar\theta_{\rm mis}=[0,\pi]$. To sit at the anthropically needed point, a small initial misalignment angle $\bar\theta_{\rm mis}$ for  $F_a> 10^{12}$ GeV may be needed. If a WIMP is the sole candidate for CDM, one obtains just one number for $\delta\rho/\rho$, and one may need a fine tuning for this to occur. The axion with $F_a>10^{12}$ GeV can choose the point anthropically. Namely, WIMPs may be dominantly the CDM, and the rest amount of CDM is provided by axions using the anthropic argument.

\section{SUSY Extension and Axino Cosmology}

The supersymmetric axion implies its superpartner {\it axino}, with a low reheating temperature. The low reheating temperature after inflation is known for a long time from the gravitino problem \cite{EKN}: $T_{\rm RH}<10^9$ GeV (old bound) or $T_{\rm RH}<10^7$ GeV (new bound if $M_{\rm gluino}<m_{3/2}$). The neutralino LSP seems the most attractive candidate for DM simply because the TeV order SUSY breaking scale introduces the LSP as a WIMP. This scenario needs an exact or an effective R-parity for proton to be sufficiently long lived. In the gauge mediated SUSY breaking scenario, the gravitino mass is generally smaller than the neutralino mass and possibly smaller than the axino mass, for which case cosmology has been studied.

\begin{figure}[!h]
\resizebox{0.5\columnwidth}{!}
{\hskip 7cm\includegraphics{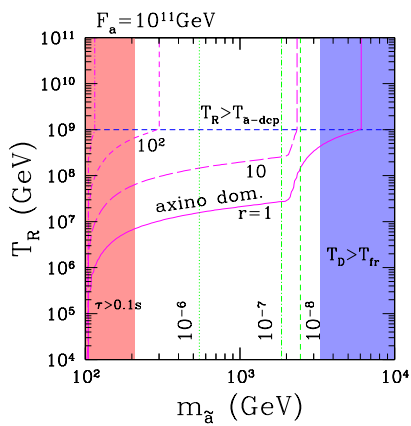}}
\caption{The heavy axino possibility \cite{ChoiKY08}. }\label{fig:Heavyaxino}
\end{figure}

There is no strong theoretical prediction on the axino mass, and we take it any value from keV to tens of TeV. For the axino lighter than the neutralino, its warm and CDM  possibilities are known for a long time \cite{ChoiKY08}. Recently, the axino heavier than the neutralino has been studied \cite{ChoiKY08}. In this case, of course the neutralino cosmology changes. If the PAMELA data \cite{PAMELAexp} is correct, this possibility of the heavy axino still survives while the earlier axino dark matter possibilities  are excluded. In Fig. \ref{fig:Heavyaxino}, we show the allowed region for the heavy axino with $F_a=10^{11}$ GeV \cite{ChoiKY08}.

\section{Conclusion}
The popular CDM candidates are WIMPs and very light axions. Direct searches for WIMPs in the universe use the WIMP cross section at Earth. The LHC machine will tell whether the LSP mass falls in the CDM needed range or not. The other candidate a very light axion, whether or not it is the dominant CDM component, is believed to exist from the need for a solution of the strong CP problem. Even though axion is not the dominant component of CDM, it may still constitute some of it.  Most exciting however would be that axion is discovered and its discovery confirms instanton physics of QCD (by experiments).


\end{document}